\documentclass[a4paper,11pt]{article}
\usepackage{pos}
\usepackage{bm}
\usepackage{times}
\usepackage{braket}
\usepackage{hyperref}
\usepackage{subfigure}
\usepackage{graphicx}
\usepackage{slashed}
\usepackage{orcidlink}
\usepackage{multirow}
\usepackage{appendix}
\usepackage{dashrule}
\usepackage{float}

\definecolor{green}{rgb}{0,0.6,0}

\newcommand{\lhc}{{\rm lhc}}
\newcommand{\rhc}{{\rm rhc}}

\newcommand{\be}{\begin{equation}} 
\newcommand{\ee}{\end{equation}}
\newcommand{\bea}{\begin{eqnarray}} 
\newcommand{\eea}{\end{eqnarray}}
\newcommand{\beas}{\begin{eqnarray*}} 
\newcommand{\eeas}{\end{eqnarray*}}

\title{$T_{cc}$ from finite volume energy levels: the left-hand cut problem and its solution	}
\author*[a]{Lu Meng}

\affiliation[a]{Institut f\"ur Theoretische Physik II, Ruhr-Universit\"at Bochum, D-44780 Bochum, Germany }

\emailAdd{lu.meng@rub.de}
\abstract{Lattice QCD has become  a crucial tool for studying hadron-hadron interactions from first principles. However, significant challenges arise when extracting infinite-volume scattering parameters from finite-volume energy levels using the conventional Lüscher method, particularly due to the presence of left-hand cuts induced by long-range interactions such as the one-pion exchange. To address these limitations, we propose a novel framework that combines chiral effective field theory and the plane-wave expansion with the Hamiltonian approach. By solving a Schrödinger-like equation in a finite volume, this method establishes a connection between finite-volume energy spectra and infinite-volume physical quantities, while effectively handling  issues caused by left-hand cuts. Furthermore, the adoption of a plane-wave basis helps mitigating complexities associated with partial-wave mixing. Our preliminary numerical results at $m_\pi \approx 280$ MeV confirm  that this approach efficiently overcomes the shortcomings of the Lüscher method and indicate a resonant interpretation of the $T_{cc}(3875)$ state—in contrast to the virtual state suggested in conventional analyses. }

\FullConference{The XVIth Quark Confinement and the Hadron Spectrum Conference (QCHSC24)\\
 19-24 August, 2024\\
 Cairns Convention Centre, Cairns, Queensland, Australia\\}

\begin{document}
\maketitle

\section{Introduction}

\begin{figure}[h]
    \centering
    \includegraphics[width=0.8\linewidth]{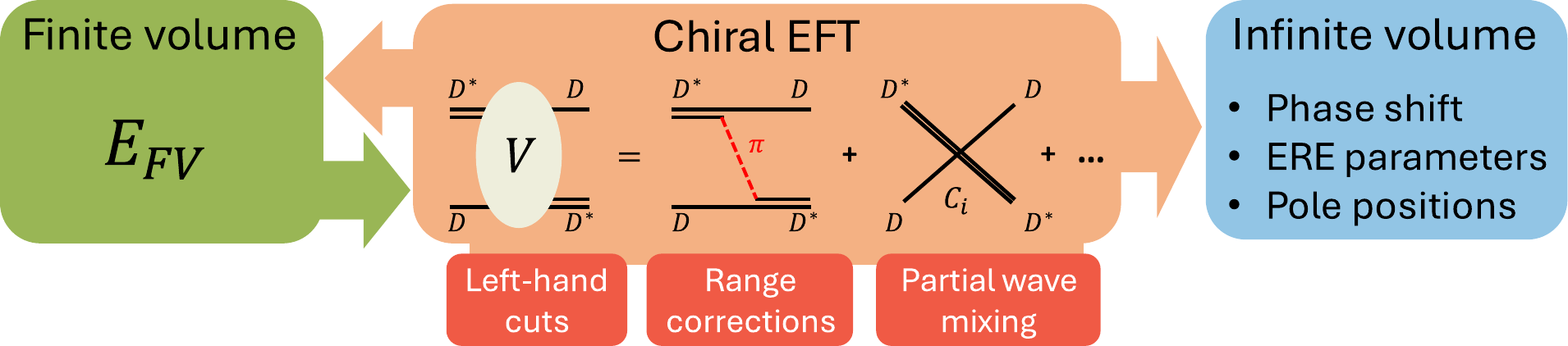}
    \caption{Schematic illustration of the approach used in this study: $ V $ represents the effective potential in chiral EFT, incorporating the one-pion exchange  and contact interactions. $ E_{FV} $ denotes the finite-volume energy levels from lattice simulations, which serve as input.}
    \label{fig:illu}
\end{figure}

In the study of hadron-hadron interactions, the primary output of lattice QCD simulations consists of discrete energy levels in a finite volume (FV). These energy levels are conventionally related to infinite-volume (IFV) scattering amplitudes through the Lüscher formalism~\cite{Luscher:1990ux}, which provides the quantization condition:  
\begin{equation}
{\rm det} \left[ G_F^{-1}(L, E) - K(E) \right] = 0. \label{eq:lusch} 
\end{equation}
Here, \( G_F \) represents the kinematic term dependent on the finite volume quantity, while \( K \) denotes the physical \(K\)-matrix for scattering. The validity of this framework requires large simulation volumes to suppress exponentially suppressed effects from long-range forces. However, in practical simulations where the box size is insufficiently large, long-range interactions become particularly problematic, leading to significant deviations. One notable issue is the left-hand cut (lhc) problem, as observed in systems like \(\Lambda\Lambda\)~\cite{Green:2021qol} and \(DD^*\)~\cite{Padmanath:2022cvl}. These problem hinder reliable extraction of scattering amplitudes from  energy levels close to the lhc. In addition, to mitigate the lhc problem and extract nucleon-nucleon interactions at the physical pion mass, a box size of \( L \gtrsim 8 \, \text{fm} \) is required~\cite{Meng:2023bmz}. Recent methodological advances \cite{Meng:2021uhz, Meng:2023bmz, Meng:2024kkp,Raposo:2023oru, Raposo:2025dkb, Bubna:2024izx, Hansen:2024ffk, Dawid:2024oey} address these limitations. Notably, an approach combining chiral effective field theory (EFT) with plane-wave expansions (Fig.~\ref{fig:illu}) has been successfully applied to analyze lattice QCD data~\cite{Meng:2021uhz, Meng:2024kkp}, with recent extensions to helicity bases~\cite{Yu:2025gzg}. This work explores its  application to for the \(DD^*\) system, highlighting its advantages over traditional methods.

In addition to the limitations caused by long-range forces, Lüscher's method faces challenges from the cubic box breaking rotational symmetry, which induces partial wave mixing in energy levels. This mixing obscures the direct relationship between individual phase shifts and finite-volume spectra, requiring a parameterized K-matrix approach for amplitude extraction. While the effective range expansion (ERE) offers one such parameterization, its applicability is severely restricted by lhcs \cite{Du:2023hlu, Du:2024snq}. Chiral EFT overcomes these problem, and when combined with the plane wave technique, additionally resolves complications from partial wave mixing.

\begin{figure}[t]
\begin{center}
\includegraphics[width=0.7\textwidth]{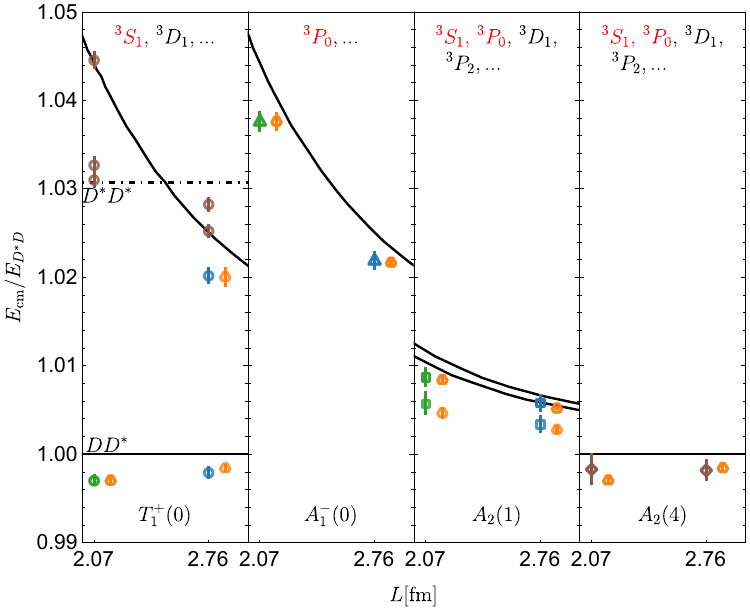}
\caption{ \label{fig:E_FV} Lattice data \cite{Padmanath:2022cvl} and fit results for the center-of-mass energy $ E_{\rm cm} = \sqrt{E^2 - {\bf P}^2} $ of the $ DD^* $ system, normalized by $ E_{DD^*} = M_D + M_{D^*} $, across various finite-volume irreps. Open circles, squares, and triangles represent the lattice energy levels, with blue and green points in the irreps $ T_1^+(0) $, $ A_1^-(0) $, and $ A_2(1) $ used as input. Orange symbols, slightly shifted to the right for clarity, represent the results of our full calculation (Fit 2), which includes pion effects. For each irrep, the contributing lowest partial waves are indicated. Predictions for the irrep $ A_2(4) $ are provided. Solid and dot-dashed lines denote the non-interacting $ DD^* $ and $ D^*D^* $ energies, respectively.}
\end{center}
\end{figure}

As an example of application, we analyze the $DD^*$ interaction using lattice data from Ref.~\cite{Padmanath:2022cvl}, where simulations were performed with $a \approx 0.08636\,\text{fm}$ at $m_{\pi} \approx 280\,\text{MeV}$. The  meson masses are $M_D = 1927\,\text{MeV}$ and $M_{D^*} = 2049\,\text{MeV}$, with the energy levels for two spatial volumes ($L = 2.07\,\text{fm}$ and $2.76\,\text{fm}$) shown in Fig.~\ref{fig:E_FV}. This system holds particular significance following the experimental discovery of the $T_{cc}(3875)$ state \cite{LHCb:2021auc,LHCb:2021vvq}, a prime tetraquark candidate with $cc\bar{u}\bar{d}$ composition located near the $DD^*$ threshold. The $T_{cc}$ provides a unique opportunity to investigate hadronic phenomena including three-body effects~\cite{Du:2021zzh}, chiral dynamics~\cite{Abolnikov:2024key,Meng:2024kkp}, and left-hand cut challenges \cite{Meng:2024kkp}, while also serving as a testbed for tetraquark stability studies \cite{Francis:2024fwf,Meng:2021jnw,Padmanath:2022cvl,Meng:2023jqk,Lyu:2023xro}.

\section{Left-hand cut problem}

\begin{figure}
    \centering
    \includegraphics[width=0.35\textwidth]{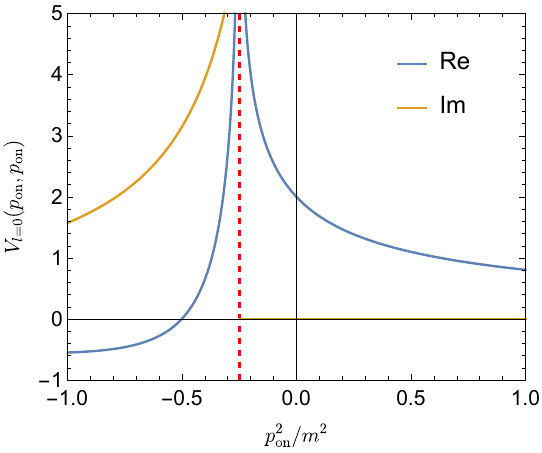}
    \caption{The S-wave on-shell potential, defined in Eq.~\eqref{eq:lhclog}. The red vertical dashed line marks the branch point of the left-hand cut at $ p_{\rm on}^{2}/m^2 = -\frac{1}{4} $. Below this point, both the potential and the corresponding $ K $-matrix become complex. The real and imaginary parts of the potential are depicted by blue and orange lines, respectively.  
    } 
    \label{fig:sm_Vlhc}
\end{figure}

To demonstrate left-hand cut effects, we consider a Yukawa potential in momentum space:
\begin{equation}
V(\bm{p},\bm{p}') = \frac{1}{(\bm{p} - \bm{p}')^2 + m^2} = \frac{1}{p^2 + p'^2 - 2pp'z + m^2},\label{eq:yukawa}
\end{equation}
where $m$ is the exchanged particle mass, $\bm{p},\bm{p}'$ denote the off-shell momenta, and $z=\hat{p}\cdot\hat{p}'$. The S-wave projection yields:
\begin{equation}
V_{l=0}(p, p') = \frac{1}{2pp'}\ln\left(\frac{(p+p')^2+m^2}{(p-p')^2+m^2}\right),~\label{eq:lhclog}
\end{equation}
revealing multivaluedness stemming from branch cuts. Using the on-shell momentum $p_{\rm on}^2=2\mu E$ (with $\mu$ as the reduced mass), we analyze singularities in the Lippmann-Schwinger equation. While $V(p,p')$ remains analytical for positive momenta, the on-shell potential develops a singularity when:
\begin{equation}
p_{\rm on}^2 < -m^2/4 \quad \text{(from $2p_{\rm on}^2(1-z)+m^2=0$)}.
\end{equation}
As shown in Fig.~\ref{fig:sm_Vlhc}, this threshold generates an imaginary component that also appears in the on-shell $K$-matrix. In the infinite volume, the cut limits the effective range expansion's convergence:
\begin{equation}
K^{-1}(p_{\rm on}) = p_{\rm on}\cot\delta(p_{\rm on}) = \frac{1}{a} + \frac{1}{2}rp_{\rm on}^2 + \cdots,\label{eq:ERE}
\end{equation}
as the left-hand cut violates the expansion's analyticity assumptions.

In the finite volume calculations, the lhc inherent to long-range interactions, fundamentally disrupts the Lüscher formalism. While the  quantity $G_F^{-1}(L,E)$ remains strictly real, the lhc induces an imaginary component in the $K$-matrix when $p_{\rm on}^2<-m^2/4$, invalidating the standard quantization condition in Eq.~\eqref{eq:lusch}. This limitation precludes a direct application of Lüscher's method to long-range interacting systems. Crucially, however, both the half off-shell potential $V_{l=0}(p_{\rm on},q)$ (for real $q>0$) and fully off-shell potentials remain analytical when $E<0$, as evident from Eq.~\eqref{eq:lhclog}. This key observation enables reliable solutions of the Schrödinger eigenvalue problem for bound states even the binding energies blew the lhc, forming the foundation of our approach. The same reasoning explains why the HAL QCD method~\cite{Aoki:2025jvi} similarly avoids lhc complications.

\section{Formalism}

The FV scattering problem is formulated through a Lippmann-Schwinger-type equation:  
\begin{equation}
\mathbb{T}(E) = \mathbb{V}(E) + \mathbb{V}(E)\mathbb{G}(E)\mathbb{T}(E).~\label{eq:LSE}
\end{equation}  
The discrete energy levels are determined by the condition:  
\begin{equation}
\det\left[\mathbb{G}^{-1}(E) - \mathbb{V}(E)\right] = 0, \label{eq:rel_det0}
\end{equation}  
where the propagator takes the form:  
\begin{equation}
\mathbb{G}_{\bm{n},\bm{n}'} = {\cal J}L^{-3}G(\tilde p_{\bm{n}},E)\delta_{\bm{n}',\bm{n}}, \quad G(\tilde p,E) = \frac{1}{4\omega_{1}\omega_{2}} \left( \frac{1}{E - \omega_1 - \omega_2} - \frac{1}{E + \omega_1 + \omega_2} \right).
\end{equation}  
Here, \(\tilde p_{\bm{n}}\) denotes the discrete momentum, and \(\mathcal{J}\) is the Jacobian factor arising from the Lorentz boost. The determinant equation can be reformulated as a Schrödinger-type eigenvalue problem, where discrete FV states emerge as bound solutions confined by the cubic box. This formulation involves only off-shell momentum, naturally avoiding the left-hand cut problem.

To construct the potential matrix in Eq.~\eqref{eq:LSE}, we use the chiral EFT potential (see Refs~\cite{Wang:2019ato,Wang:2019nvm,Wang:2020dko,Meng:2022ozq}. for general applications of the Chiral EFT),, which combines short-range contact terms with long-range one-pion exchange (OPE) contributions:  
\begin{equation}
V = V_{\text{OPE}}^{(0)} + V_{\text{cont}}^{(0)} + V_{\text{cont}}^{(2)} + \dots, \label{eq:veft}
\end{equation}  
where \( Q \sim m_{\pi} \) characterizes the small scale of the expansion. To \(\mathcal{O}(Q^2)\), the relevant contact interactions are given by:  
\begin{equation}
\begin{aligned}
V_{\text{cont}}^{(0)+(2)}[^3S_1] &= \left(C^{(0)}_{^3S_1} + C^{(2)}_{^3S_1}(p^2 + p'^2)\right) (\bm{\epsilon} \cdot \bm{\epsilon}'^*), \\
V_{\text{cont}}^{(2)}[^3P_0] &= C^{(2)}_{^3P_0} (\bm{p}' \cdot \bm{\epsilon}'^*)(\bm{p} \cdot \bm{\epsilon}). \label{eq:Vct}
\end{aligned}
\end{equation}  
Here, \(\bm{p}^{(\prime)}\) and \(\bm{\epsilon}^{(\prime)}\) denote the momenta and polarizations of \(D^{(*)}\) mesons, respectively. Additionally, the static OPE potential is expressed as:  
\begin{equation}
V_{\text{OPE}}^{(0)} = -3\frac{M_D M_{D^*} g^2}{f_{\pi}^2} \frac{(\bm{k} \cdot \bm{\epsilon})(\bm{k} \cdot \bm{\epsilon}'^*)}{\bm{k}^2 + \mu^2}, \label{Eq:OPE}
\end{equation}  
where \(\mu^2 = m_{\pi}^2 - \Delta M^2\) (with \(\Delta M \equiv M_{D^*} - M_D\)) and \(\bm{k} = \bm{p}' + \bm{p}\). Using \(f_{\pi} = 105.3 \, \text{MeV}\)~\cite{Du:2023hlu,Becirevic:2012pf} and \(g = 0.517(15)\) for \(a \approx 0.086 \, \text{fm}\), the leading lhc branch point is found to be:  
\begin{equation}
(p_{\lhc}^{1\pi})^2 = -\mu^2/4 = -(126 \, \text{MeV})^2, \quad \left(\frac{p_{\lhc}^{1\pi}}{E_{DD^*}}\right)^2 \approx -0.001. \label{eq:lhc_num}
\end{equation}  
The three-body \(DD\pi\) threshold (\(p_{\rhc_3}^2 = (552 \, \text{MeV})^2\)) lies beyond the energy range considered here. Crucially, the plane-wave method employed in this work preserves all partial waves in the OPE potential without truncation.  Two-pion exchange contributions at the considered $m_{\pi}$ value are assumed to be absorbed into the contact terms (see \cite{Chacko:2024ypl}

We regulate the contact terms in the Lippmann-Schwinger equation with the regulator:
\begin{equation}
e^{-(p^n + p'^n)/\Lambda^n} \quad (n=6),
\end{equation}
while preserving long-range dynamics through modified pion propagators:
\begin{equation}
\frac{1}{\bm{k}^2 + \mu^2} \rightarrow \frac{1}{\bm{k}^2 + \mu^2} e^{-(\bm{k}^2+\mu^2)/\Lambda^2},
\end{equation}
following Ref.~\cite{Reinert:2017usi}. Using $\Lambda$= 0.9 GeV, we find cutoff variations induce negligible uncertainties~\cite{Meng:2023bmz}. The low-energy constants are constrained by fits to $m_\pi$ = 280 MeV lattice data.

\section{Numerical results}

\begin{figure}
    \centering
    \includegraphics[width=0.95\linewidth]{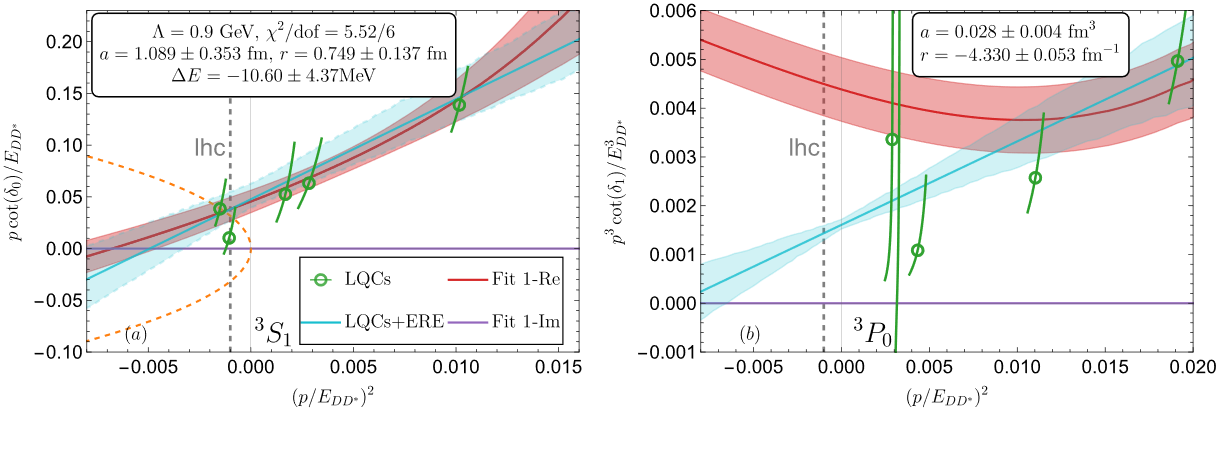}
      \includegraphics[width=0.95\linewidth]{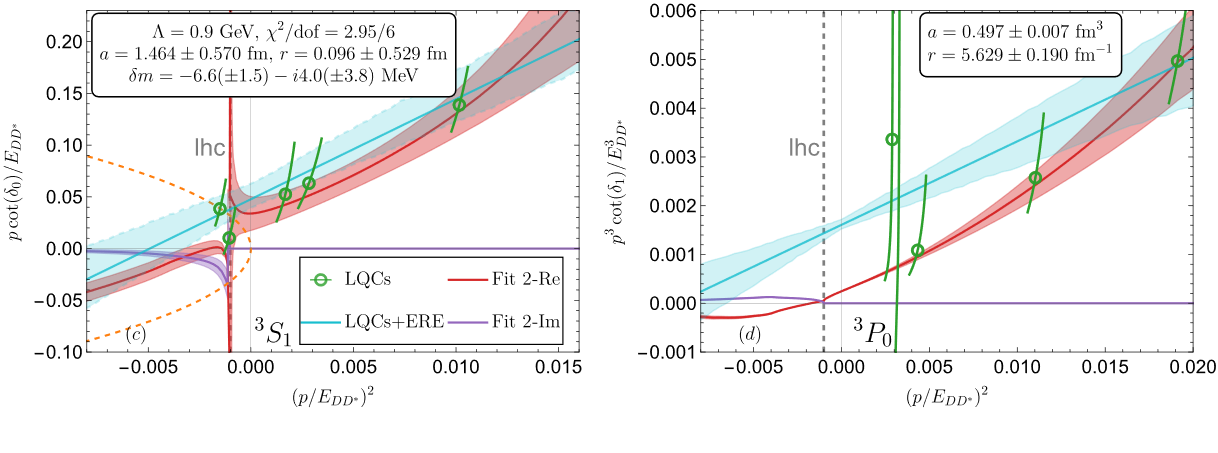}
    \caption{Phase shifts in the ${}^3S_1$ (left panel) and ${}^3P_0$ (right panel) partial waves extracted from lattice QCD data. Red bands represent the results of our 3-parameter fits without the OPE (Fit1, upper  panel) and with the OPE (Fit2, lower panel),  including the $1\sigma$ uncertainty. Green dots in the left panel are the phase shifts using the single-channel L\"uscher quantization conditions in Ref.~\cite{Padmanath:2022cvl}. Green dots in the right panel are extracted in this study using the same method. Blue bands are the results of the 4-parameter fits using the ERE in
Ref.~\cite{Padmanath:2022cvl}. Orange lines in the left panel correspond to $ip=\pm |p|$ from unitarity, normalized to $E_{DD^*}$. The gay vertical dashed line denotes the position of the branch point of the left-hand cut nearest to the threshold.}
    \label{fig:phase_shift}
\end{figure}

We perform two analyses to evaluate the effects of one-pion exchange (OPE): Fit 1, which uses a pure contact potential with optimized low-energy constants (LECs), and Fit 2, which incorporates both contact and OPE potentials. The energy level results are presented in Fig.~\ref{fig:E_FV}, while the phase shifts, effective range parameters, and pole information for the \( ^3S_1 \) and \( ^3P_0 \) partial waves are shown in Fig.~\ref{fig:phase_shift}.

For \(\delta_{^3S_1}\) (upper left panel of Fig.~\ref{fig:phase_shift}), the predictions from Fit 1 align closely with the ERE analysis in Ref.~\cite{Padmanath:2022cvl} (Eq.~\ref{eq:ERE}), yielding comparable parameters and a similar \(T_{cc}\) pole position. This agreement is expected, as both approaches use two parameters to match the scattering length and effective range. However, the contact-only fit struggles to describe the \(\delta_{^3P_0}\) data, primarily due to the limited number of parameters available to capture the low-energy behavior accurately.  In contrast, while Ref.~\cite{Padmanath:2022cvl} addressed range effects by introducing an additional ERE parameter, our analysis shows this is unnecessary. The inclusion of the OPE in Fit 2 naturally accounts for these corrections, as demonstrated in the lower right panel. Furthermore, the OPE significantly impacts the dynamics of \(\delta_{^3S_1}\): the interplay between repulsive pion exchange and attractive short-range interactions generates a pole in \(p \cot \delta_{^3S_1}\) near the lhc, consistent with the findings in Ref.~\cite{Du:2023hlu}. The extracted \(T_{cc}\) pole position suggests the presence of a resonance state rather than a virtual state.  Comparisons between the Lüscher-method created phase shifts (green points) and Fit 2 reveal deviations at low energies, particularly for two data points strongly influenced by the lhc. However, above the \(DD^*\) threshold, the results for both \(\delta_{^3S_1}\) and \(\delta_{^3P_0}\) become consistent within uncertainties.

\section{Conclusion}

We introduce a novel EFT-based method to extract two-body scattering observables from finite-volume energies. In contrast to the Lüscher approach, our framework explicitly includes the long-range interaction and the leading left-hand cut, ensuring the correct analytic structure of the scattering amplitude near threshold. By computing finite-volume energy levels as eigenvalue solutions—both below and above the left-hand cut—we systematically account for range effects and leading exponential corrections in a model-independent way.

By employing a plane wave basis expansion, our method efficiently accounts for partial-wave mixing effects. We show the validity of this approach by performing a detailed study of $DD^*$ system~\cite{Padmanath:2022cvl}, which is crucial for understanding the doubly charmed tetraquark. Notably, the OPE-induced long-range interaction plays a pivotal role in infinite-volume observables, particularly in the $^3P_0$ channel, where it resolves left-hand cut limitations inherent to the Lüscher formalism. EFT truncation effects prove negligible relative to statistical errors, reinforcing the robustness of our results.  Our findings suggest that the $T_{cc}^+$ state is a below-threshold resonance. Future improvements in lattice precision will allow a direct extraction of the OPE coupling $g/f_{\pi}$ from data. Additionally, our framework could be extended to  accommodate for the three-body ($DD\pi$) right-hand cut~\cite{Hansen:2024ffk}.

A similar EFT-based analysis of the isovector $DD^*$ scattering lattice data at $m_{\pi} = 280$ MeV is presented in Ref.~\cite{Meng:2024kkp}. Our method has wide applicability across hadronic systems at unphysical pion masses, particularly for cases where finite-volume energy levels exist or are anticipated from lattice QCD, including nucleon-nucleon, hyperon-nucleon, hyperon-hyperon, tetraquarks, pentaquarks and six-quark states.

\section*{Acknowledgment}
This work has been supported in part by the European Research Council (ERC AdG NuclearTheory, grant No.885150).

% \begin{thebibliography}{99}
% \bibitem{...}
% ....

% \end{thebibliography}

\bibliographystyle{JHEP}
\bibliography{ref.bib}

\end{document}